\documentclass[aps,preprint,showpacs, showkeys]{revtex4}%
\usepackage{amsfonts}
\usepackage{amsmath}
\usepackage{amssymb}
\usepackage{graphicx}%
\providecommand{\U}[1]{\protect\rule{.1in}{.1in}}

\begin{document}
\title[ ]{New Ansatz for Metric Operator Calculation in Pseudo-Hermitian Field Theory}
\author{Abouzeid M. Shalaby}
\email{amshalab@ mans.edu.eg}
\affiliation{Physics Department, Faculty of Science, Mansoura University, Egypt.}

\begin{abstract}
In this work, a new  ansatz is introduced to make the calculations of the
metric operator in Pseudo-Hermitian field theory simpler. The idea is to
assume that the metric operator is not only a functional of the field
operators $\phi$ and its conjugate field $\pi$ but also on the field gradient
$\nabla\phi$. Rather than the locality of the metric operator obtained, the
ansatz enables one to calculate the metric operator just once for all
dimensions of the space-time. We calculated the metric operator of the
$i\phi^{3}$ scalar field theory up to first order in the coupling. The higher
orders can be conjectured from their corresponding operators in the quantum
mechanical case available in the literature. We assert that, the calculations
existing in literature for the metric operator in field theory are cumbersome
and are done case by case concerning the dimension of space-time in which the
theory is investigated. Moreover, while the resulted metric operator in this
work is local, the existing calculations for the metric operator leads to a
non-local one. Indeed, we expect that the new results introduced in this work
will greatly lead to the progress of the studies in Pseudo-Hermitian field
theories where there exist a lack of such kind of studies in the literature. In fact, with the aid of this work a rigorous study of a $\mathcal{PT}$-symmetric Higgs mechanism can be reached.

\end{abstract}
\maketitle

The subject of quantum mechanical pseudo-Hermitian models has attained
extensive attention in the literature
\cite{aboeff,bendg1,Conj,bend1,bend2,bend3,bend4,bend5,bend6,bend7,bend8,bend9,bend10,bend11,bend12,bend13,bend14}%
. Two main problems have appeared at the beginning of these studies, namely,
the indefinite norm and unitarity problems. In fact, there exists two equivalent
approaches to solve these problems. The first approach, due to Bender
$et.\ al$, is to replace the bra in the Dirac convention by a $CPT$ bra such
that the new inner product takes the form \cite{bend2005}
\[
\langle A|B\rangle_{CPT}=(CPT|A\rangle)^{T}|B\rangle,
\]
which replaces the conventional Dirac bracket $\ \langle A|B\rangle
=|A\rangle^{\dagger}|B\rangle.$ The operator $C$ is determined dynamically and
in most cases perturbatively. This approach succeeded in correcting  the
negative norm  and unitarity problems.

The other approach, due to  Mostafazadeh \cite{zadah}, has generalized the
requirement of a non-Hermitian theory to have a real spectrum to the existence
of a positive definite metric operator $\eta$ such that $H^{\dagger}=\eta
H\eta^{-1}$. The existence of the metric operator assures the existence of a
similarity transformation which has the job of transforming the non-Hermitian
Hamiltonian operator $H$ into a Hermitian operator $h$, where $h=\rho
^{-1}H\rho$ and $\rho=\sqrt{\eta}$.

In both approaches described above, the metric operator $\eta$ and the $C$
operator are assumed to be composed of the dynamical operators (momentum and
position operators) in the Hamiltonian. While these approaches can be applied
in both the quantum mechanical and quantum field versions of \ a Hamiltonian,
their application in field theory leads to complicated $\eta$ and $C$
operators \cite{cop}. This is the reason why in the literature one can find even up to
seventh order calculation of $C$ for the $ix^{3}$ while only a first order at
low space-time dimensions calculation of the same operator exists for the
quantum field $i\phi^{3}$ scalar theory. In this work, we introduce a new
ansatz that makes the calculation of the metric operator in field theory goes
simple as in the quantum mechanical case. The idea is that, instead of
conjecturing the shape of the metric operator as a functional of the field
$\phi$ and its conjugate momentum field $\pi$, we assume that it can also
include the field gradient $\nabla\phi$. As we will see this assumption will
simplify the calculation and lead to metric operator with parameters that do
not depend on the dimension of space-time. Note that the metric operator is not
unique and one can find more than one metric operator for the same theory.

To start, consider the Hamiltonian model of the form;%
\begin{equation}
H=\frac{1}{2}\left(  \left(  \nabla\phi\right)  ^{2}+\pi^{2}+m^{2}\phi
^{2}\right)  +ig\phi^{3}, \label{ham}%
\end{equation}
where $\phi$ is the field operator, $\pi$ is the conjugate momentum and $g$ is
the coupling constant. This theory is non-Hermitian but $\mathcal{PT}%
$-symmetric and thus is physically acceptable as it has a real spectrum. Also,
following Mostafazadeh, the metric operator can be assumed to have the form
$\eta=\exp\left(  -Q\right)  $, where $Q$ is Hermitian. Note that, although
the Hamiltonian in Eq.(\ref{ham}) is non-Hermitian in a Hilbert space with the
Dirac sense inner product, it is Hermitian in a Hilbert space endowed by the
inner product $\langle n|\eta|n\rangle$.

To obtain $Q$, we rename the terms in the Hamiltonian as;
\[
H=H_{0}+gH_{I}%
\]%
\begin{align*}
H_{0}  &  =\frac{1}{2}\int d^{3}x\left(  \left(  \nabla\phi\right)  ^{2}%
+\pi^{2}+m^{2}\phi^{2}\right)  ,\\
H_{I}  &  =i\int d^{3}x\phi^{3},
\end{align*}
and in using the relation $H^{\dagger}=\eta H\eta^{-1}$, we get%

\begin{align*}
H^{\dagger}  &  =\exp(-Q)H\exp(Q)=H+[-Q,H]+\frac{[-Q,[-Q,H]]}{2!}\\
&  +\frac{[-Q,[-Q,[-Q,H]]]}{3!}+....,
\end{align*}
where%
\[
Q=Q_{0}+gQ_{1}+g^{2}Q_{2}++g^{3}Q_{3}+..
\]
Accordingly;%
\begin{align*}
-2gH_{I}  &  =\left[  -\left(  Q_{0}+gQ_{1}+g^{2}Q_{2}++g^{3}Q_{3}+...\right)
,H\right] \\
&  +\frac{%
\begin{array}
[c]{c}%
\lbrack-\left(  Q_{0}+gQ_{1}+g^{2}Q_{2}++g^{3}Q_{3}+...\right)  ,\\
\lbrack-\left(  Q_{0}+gQ_{1}+g^{2}Q_{2}++g^{3}Q_{3}+...\right)  ,H]]
\end{array}
}{2!}\\
&  +\frac{\left[
\begin{array}
[c]{c}%
\lbrack-\left(  Q_{0}+gQ_{1}+g^{2}Q_{2}++g^{3}Q_{3}+...\right)  ,\\
\lbrack-\left(  Q_{0}+gQ_{1}+g^{2}Q_{2}++g^{3}Q_{3}+...\right)  ,\\
\lbrack-\left(  Q_{0}+gQ_{1}+g^{2}Q_{2}++g^{3}Q_{3}+...\right)  ,H]]
\end{array}
\right]  }{3!}\\
&  +................................
\end{align*}
Then by equating coefficients of $g^{n}$ in each side, we get the first seven
orders as follows \footnote{ This set of equations have been obtained in Ref.\cite{cop} but they employed the results from each equation in the next one.};
\begin{align*}
g^{1}  &  :-2H_{I}=\left[  -Q_{1},H_{0}\right]  ,\\
g^{3}  &  :0=\left[  -Q_{3},H_{0}\right]  +\frac{\left[  -Q_{1},\left[
-Q_{1},\left[  -Q_{1},H_{0}\right]  \right]  \right]  }{3!}+\frac{\left[
-Q_{1},\left[  -Q_{1},H_{I}\right]  \right]  }{2!}\\
g^{5}  &  :0=\left[  -Q_{5},H_{0}\right]  +\frac{\left[  -Q_{1},\left[
-Q_{1},\left[  -Q_{1},\left[  -Q_{1},\left[  -Q_{1},H_{0}\right]  \right]
\right]  \right]  \right]  }{5!}+\frac{\left[  -Q_{3},\left[  -Q_{1},\left[
-Q_{1},H_{0}\right]  \right]  \right]  }{3!}\\
&  +\frac{\left[  -Q_{1},\left[  -Q_{3},\left[  -Q_{1},H_{0}\right]  \right]
\right]  }{3!}+\frac{\left[  -Q_{1},\left[  -Q_{1},\left[  -Q_{3}%
,H_{0}\right]  \right]  \right]  }{3!}\\
&  +\frac{\left[  -Q_{1},\left[  -Q_{1},\left[  -Q_{1},\left[  -Q_{1}%
,H_{I}\right]  \right]  \right]  \right]  }{4!}+\frac{\left[  -Q_{1},\left[
-Q_{3},H_{I}\right]  \right]  }{2!}+\frac{\left[  -Q_{3},\left[  -Q_{1}%
,H_{I}\right]  \right]  }{2!},\\
g^{7}  &  :0=\left[  -Q_{7},H_{0}\right]  +\frac{\left[  -Q_{1},\left[
-Q_{1},\left[  -Q_{1},\left[  -Q_{1},\left[  -Q_{1},\left[  -Q_{1},\left[
-Q_{1},H_{0}\right]  \right]  \right]  \right]  \right]  \right]  \right]
}{7!}\\
&  +\frac{\left[  -Q_{1},\left[  -Q_{1},\left[  -Q_{1},\left[  -Q_{1},\left[
-Q_{3},H_{0}\right]  \right]  \right]  \right]  \right]  }{5!}\\
&  +\frac{\left[  -Q_{1},\left[  -Q_{1},\left[  -Q_{1},\left[  -Q_{3},\left[
-Q_{1},H_{0}\right]  \right]  \right]  \right]  \right]  }{5!}+\frac{\left[
-Q_{1},\left[  -Q_{1},\left[  -Q_{3},\left[  -Q_{1},\left[  -Q_{1}%
,H_{0}\right]  \right]  \right]  \right]  \right]  }{5!}\\
&  +\frac{\left[  -Q_{1},\left[  -Q_{3},\left[  -Q_{1},\left[  -Q_{1},\left[
-Q_{1},H_{0}\right]  \right]  \right]  \right]  \right]  }{5!}+\frac{\left[
-Q_{3},\left[  -Q_{1},\left[  -Q_{1},\left[  -Q_{1},\left[  -Q_{1}%
,H_{0}\right]  \right]  \right]  \right]  \right]  }{5!}\\
&  +\frac{\left[  -Q_{3},\left[  -Q_{3},\left[  -Q_{1},H_{0}\right]  \right]
\right]  }{3!}+\frac{\left[  -Q_{3},\left[  -Q_{1},\left[  -Q_{3}%
,H_{0}\right]  \right]  \right]  }{3!}+\frac{\left[  -Q_{1},\left[
-Q_{3},\left[  -Q_{3},H_{0}\right]  \right]  \right]  }{3!}\\
&  +\frac{\left[  -Q_{1},\left[  -Q_{1},\left[  -Q_{5},H_{0}\right]  \right]
\right]  }{3!}+\frac{\left[  -Q_{1},\left[  -Q_{5},\left[  -Q_{1}%
,H_{0}\right]  \right]  \right]  }{3!}+\frac{\left[  -Q_{5},\left[
-Q_{1},\left[  -Q_{5},H_{0}\right]  \right]  \right]  }{3!}\\
&  +\frac{\left[  -Q_{1},\left[  -Q_{1},\left[  -Q_{1},\left[  -Q_{1},\left[
-Q_{1},\left[  -Q_{1},H_{I}\right]  \right]  \right]  \right]  \right]
\right]  }{6!}+\frac{\left[  -Q_{3},\left[  -Q_{3},H_{I}\right]  \right]
}{2!}\\
&  +\frac{\left[  -Q_{1},\left[  -Q_{1},\left[  -Q_{1},\left[  -Q_{3}%
,H_{I}\right]  \right]  \right]  \right]  }{4!}+\frac{\left[  -Q_{1},\left[
-Q_{1},\left[  -Q_{3},\left[  -Q_{1},H_{I}\right]  \right]  \right]  \right]
}{4!}\\
&  +\frac{\left[  -Q_{1},\left[  -Q_{3},\left[  -Q_{1},\left[  -Q_{1}%
,H_{I}\right]  \right]  \right]  \right]  }{4!}+\frac{\left[  -Q3,\left[
-Q_{1},\left[  -Q_{1},\left[  -Q_{1},H_{I}\right]  \right]  \right]  \right]
}{4!}+\frac{\left[  -Q_{5},\left[  -Q_{1},H_{I}\right]  \right]  }{2!}\\
&  +\frac{\left[  -Q_{1},\left[  -Q_{5},H_{I}\right]  \right]  }{2!}.
\end{align*}

In the literature, the $Q$ operator is assumed to depend only on the position
and momentum operators \cite{cop}. This assumption has been mapped to quantum
field problems with the $Q$ operator is a functional in the field $\phi$ and
its canonical conjugate field $\pi$. However, there exists an operator in the
Hamiltonian ($\nabla\phi$) which has no analog in the quantum mechanical case.
Accordingly, it makes sense to extend the assumption for the $Q$ operator to
be a functional of $\phi$, $\pi$ and $\nabla\phi$ fields. Accordingly, one can
conjecture the form of $Q_{1}$ to be;
\begin{align}
Q_{1}  &  =C_{1}\int d^{3}z\pi^{3}(z)+\frac{C_{2}}{3}\int d^{3}z\left(
\pi(z)\phi^{2}(z)+\phi(z)\pi(z)\phi(z)+\phi^{2}(z)\pi(z)\right) \nonumber\\
&  +\frac{C_{3}}{3}\int d^{3}z\left(  \pi(z)\nabla\phi(z)\nabla\phi
(z)+\nabla\phi(z)\pi(z)\nabla\phi(z)+\nabla\phi(z)\nabla\phi(z)\pi(z)\right)
.
\end{align}

To find out the parameters $C_{1}$, $C_{2}$ and $C_{3}$ we consider the first
equation in the above set;%
\[
-2H_{I}=\left[  -Q_{1},H_{0}\right]  .
\]
Let us commute each term in $Q_{1}$ with each term in $H_{0}$.

First, consider the commutator of the first term in $Q_{1}$ with each term in
$H_{0}$ ;
\begin{align*}
&  C_{1}\int d^{3}x\int d^{3}z\left[  \pi^{3}(z),\frac{1}{2}\nabla_{x}%
\phi\left(  x\right)  \nabla_{x}\phi\left(  x\right)  \right]  ,\\
&  =\frac{1}{2}C_{1}\int d^{3}x\int d^{3}z\left[  \pi^{3}(z),\nabla_{x}%
\phi\left(  x\right)  \right]  \nabla_{x}\phi\left(  x\right) \\
&  +\frac{1}{2}C_{1}\int d^{3}x\int d^{3}z\nabla_{x}\phi\left(  x\right)
\left[  \pi^{3}(z),\nabla_{x}\phi\left(  x\right)  \right]  ,\\
&  =\frac{1}{2}C_{1}\int d^{3}x\int d^{3}z\nabla_{x}\left[  \pi^{3}%
(z),\phi\left(  x\right)  \right]  \nabla_{x}\phi\left(  x\right) \\
&  +\frac{1}{2}C_{1}\int d^{3}x\int d^{3}z\nabla_{x}\phi\left(  x\right)
\nabla_{x}\left[  \pi^{3}(z),\phi\left(  x\right)  \right]  ,\\
&  =\frac{3}{2}C_{1}\int d^{3}x\int d^{3}z\pi^{2}(z)\nabla_{x}(-i\delta
^{3}(z-x)\nabla_{x}\phi\left(  x\right)  ,\\
&  +\frac{3}{2}C_{1}\int d^{3}x\int d^{3}z\nabla_{x}\phi\left(  x\right)
\pi^{2}(z)\nabla_{x}(-i\delta^{3}(z-x),\\
&  =\frac{3}{2}C_{1}\int d^{3}x\int d^{3}z\pi^{2}(z)\nabla_{x}(i\delta
^{3}(x-z)\nabla_{x}\phi\left(  x\right)  ,\\
&  +\frac{3}{2}C_{1}\int d^{3}x\int d^{3}z\nabla_{x}\phi\left(  x\right)
\pi^{2}(z)\nabla_{x}(i\delta^{3}(x-z),\\
&  =-\frac{3i}{2}C_{1}\int d^{3}x\pi^{2}(x)\nabla_{x}^{2}\phi\left(  x\right)
-\frac{3i}{2}C_{1}\int d^{3}x\nabla_{x}^{2}\phi\left(  x\right)  \pi^{2}(x).
\end{align*}
Also%
\begin{align}
&  \frac{1}{2}m^{2}C_{1}\int d^{3}x\int d^{3}z\left[  \pi^{3}(z),\phi\left(
x\right)  \phi\left(  x\right)  \right]  ,\nonumber\\
&  =\frac{1}{2}m^{2}C_{1}\int d^{3}x\int d^{3}z\left[  \pi^{3}(z),\phi\left(
x\right)  \right]  \phi\left(  x\right) \nonumber\\
&  +\frac{1}{2}m^{2}C_{1}\int d^{3}x\int d^{3}z\phi\left(  x\right)  \left[
\pi^{3}(z),\phi\left(  x\right)  \right]  ,\nonumber\\
&  =\frac{3}{2}m^{2}C_{1}\int d^{3}x\int d^{3}z\pi^{2}(z)\left(  -i\delta
^{3}(x-z\right)  \phi\left(  x\right) \label{T2}\\
&  +\frac{3}{2}m^{2}C_{1}\int d^{3}x\int d^{3}z\phi\left(  x\right)  \pi
^{2}(z)\left(  -i\delta^{3}(x-z\right)  ,\nonumber\\
&  =\frac{-3i}{2}m^{2}C_{1}\int d^{3}x\left(  \pi^{2}(x)\phi\left(  x\right)
+\phi\left(  x\right)  \pi^{2}(x)\right)  .\nonumber
\end{align}

For the second term in $Q_{1}$ with each term in $H_{0}$, we get

\begin{align}
&  \frac{C_{2}}{6}\int d^{3}x\int d^{3}z\left[  \pi(z)\phi^{2}(z)+\phi
(z)\pi(z)\phi(z)+\phi^{2}(z)\pi(z),\pi^{2}(x)\right]  ,\nonumber\\
&  =iC_{2}\int d^{3}x\left(  \phi(x)\pi^{2}(x)+\pi^{2}(x)\phi(x)\right)  ,
\end{align}
Now%

\begin{align*}
&  \frac{C_{2}}{6}\int d^{3}x\int d^{3}z\left[  \pi(z)\phi^{2}(z),\nabla
_{x}\phi\left(  x\right)  \nabla_{x}\phi\left(  x\right)  \right]  ,\\
&  =\frac{C_{2}}{6}\int d^{3}x\int d^{3}z\nabla_{x}\left(  -i\delta
^{3}(z-x)\right)  \nabla_{x}\phi\left(  x\right)  \phi^{2}(z),\\
&  +\frac{C_{2}}{6}\int d^{3}x\int d^{3}z\nabla_{x}\phi\nabla_{x}\left(
-i\delta^{3}(z-x)\right)  \left(  x\right)  \phi^{2}(z),\\
&  =-\frac{iC_{2}}{3}\int d^{3}x\nabla_{x}^{2}\phi\left(  x\right)  \phi
^{2}(x).
\end{align*}
Then
\begin{align*}
&  \frac{C_{2}}{6}\int d^{3}x\int d^{3}z\left[  \pi(z)\phi^{2}(z)+\phi
(z)\pi(z)\phi(z)+\phi^{2}(z)\pi(z),\nabla_{x}\phi\left(  x\right)  \nabla
_{x}\phi\left(  x\right)  \right]  ,\\
&  =-iC_{2}\int d^{3}x\nabla_{x}^{2}\phi\left(  x\right)  \phi^{2}(x),
\end{align*}

and
\begin{align*}
&  \frac{C_{2}m^{2}}{6}\int d^{3}x\int d^{3}z\left[  \pi(z)\phi^{2}%
(z)+\phi(z)\pi(z)\phi(z)+\phi^{2}(z)\pi(z),\phi\left(  x\right)  \phi\left(
x\right)  \right]  ,\\
&  =-iC_{2}m^{2}\int d^{3}x\phi^{3}\left(  x\right)  .
\end{align*}

For the third term of $Q_{1}$ with each term in $H_{0};$%
\begin{align*}
&  \frac{C_{3}}{6}\int d^{3}x\int d^{3}z\left[  \nabla_{z}\phi(z)\nabla
_{z}\phi(z)\pi(z)+\nabla_{z}\phi(z)\pi(z)\nabla_{z}\phi(z)+\pi(z)\nabla
_{z}\phi(z)\nabla_{z}\phi(z),\pi^{2}(x)\right]  ,\\
&  =\frac{iC_{3}}{2}\int d^{3}x\pi(x)\pi(x)\nabla^{2}\phi(x)+\frac{iC_{3}}%
{2}\int d^{3}x\nabla^{2}\phi(x)\pi(x)\pi(x).
\end{align*}

Now consider;%
\begin{align*}
\frac{C_{3}}{6}\int d^{3}x\int d^{3}z\left[  \pi(z)\nabla\phi(z)\nabla
\phi(z),\nabla_{x}\phi(x)\nabla_{x}\phi(x)\right]   &  =\frac{2iC_{3}}{6}\int
d^{3}x\nabla\frac{\left(  \nabla\phi(x)\right)  ^{3}}{3},\\
&  =\frac{2iC_{3}}{6}\int d\frac{\left(  \nabla\phi(x)\right)  ^{3}}{3},
\end{align*}

which is an integration of a total derivative that can be integrated out and
thus vanish.

Also,%
\begin{align*}
&  \frac{m^{2}C_{3}}{6}\int d^{3}x\int d^{3}z\left[
\begin{array}
[c]{c}%
\nabla_{z}\phi(z)\nabla_{z}\phi(z)\pi(z)+\nabla_{z}\phi(z)\pi(z)\nabla_{z}%
\phi(z)\\
+\pi(z)\nabla_{z}\phi(z)\nabla_{z}\phi(z),\phi(x)\phi(x)
\end{array}
\right]  ,\\
&  =\frac{im^{2}C_{3}}{2}\int d^{3}x\phi^{2}(x)\nabla^{2}\phi(x).
\end{align*}
Note that, the calculations above used integration by parts as well as
employed the different properties of the derivative of the Dirac delta function.

Now for the relation $\left[  Q_{1},H_{0}\right]  =2H_{I}$ to be verified, one
have to take
\[
C_{2}=\frac{-2}{m^{2}}\text{, \ }C_{1}=-\frac{4}{3}\frac{1}{m^{4}}\text{,
}C_{3}=\frac{-4}{m^{4}}.
\]
To realize how advantageous is our ansatz used in this work, compare the
metric operator obtained here with that obtained in Ref. \cite{cop}. To shed
light on the differences between the two results we note that although the
algorithm used in Ref.\cite{cop} can applied to the field theory in
Eq.(\ref{ham}) for the calculation of the $Q$ operator, the results their have
parameters which depend on the space-time dimensions  while our result is general and can be used for any space-time
dimensions. Moreover, while the from in Ref.\cite{cop} is complicated and non-local, the form we obtained is local and simple and thus lead to simpler calculations of the Physical amplitudes. Besides, the modification we add here
to the form of the $Q$ operator makes the higher orders calculations in
quantum field problems as simple as those in the quantum mechanical case and
can, in fact, be conjectured from the corresponding results in the literature
for the $ix^{3}$ model. However, the higher order calculations as well as
detailed analysis of the theory under consideration is postponed to appear in
another article.

To conclude, we have introduced a new ansatz for the metric operator
calculation in \ quantum field theory. Without the ansatz introduced in this
work, the metric operator in field theory is cumbersome and non-local. 

We applied the ansatz for the $i\phi^{3}$ scalar field theory for which we
obtained the metric operator up to first order in the coupling constant $g$.
We realized that the parameters appearing in the form of the metric operator
are just the original parameters in the Hamiltonian model which are
independent on the dimension of the space-time. This result is very important
as for the current algorithms in the literature one has to do the calculations
for a fixed space-time dimensions. Moreover, the previous calculations of the
metric operator resulted in a very complicated form which in turn will make the calculations of the physical amplitudes very. On the other hand our ansatz results in a shape that
is very similar to the quantum mechanical shape of the metric operator and
though we have done the calculations up to first order, one can conjecture the
higher orders from the available metric operator in $0+1$ dimensions (quantum
mechanics). The importance of this work stems from the fact that the effective field approach of the more important $-\phi^4$ has been proved to be successful in studying this theory \cite{aboeff}. In fact, the effective from of the $-\phi^4$ is real-line theory with an $i\phi^3$ non-Hermitian term and thus with the aid of this work a serious study of the $\mathcal{PT}$-symmetric Higgs mechanism is now available.  
 
\begin{acknowledgments}
The author would like to thank Dr. ~S.~A. Elwakil for his and kind help.  Deep thanks to  the Physics department in Qassim university for support and  help.
\end{acknowledgments}

\end{document}